\documentclass[11pt, preprint2]{aastex}
\usepackage[]{natbib}

\begin{document}

\title{Evidence that the Bursting Component of the X-ray Radiation From 3C 111 Originates in the PC-Scale Jet}

\author{M.B. Bell\altaffilmark{1}
and S.P. Comeau\altaffilmark{1}}

\altaffiltext{1}{Herzberg Institute of Astrophysics,
National Research Council of Canada, 100 Sussex Drive, Ottawa,
ON, Canada K1A 0R6;
morley.bell@nrc-cnrc.gc.ca}

\begin{abstract}

Evidence is presented indicating that the bursting component of the X-ray radiation detected in the nuclear region of the active radio galaxy 3C 111 comes from the blobs ejected in the pc-scale jet and not from the accretion disc. After each new outburst the radio flux density associated with it increases to a peak in $\sim1$ year and then subsides over a period of 1-2 years with the flux falling off exponentially as the blob moves outward and dissipates. Similar peaks (bursts) are seen in the X-ray light curve and a cross-correlation between the two shows a very high correlation with the X-ray peaks leading the radio peaks by $\sim100$ days. A second cross-correlation, this time between the radio event start times and the X-ray light curve, also shows a significant correlation. When this is taken together with the long ($\sim1$ yr) delay between the start of each ejection event and its associated X-ray peak it indicates that this bursting component of the X-ray flux must be associated with the ejected blobs in the pc-scale jet and not with the accretion disc. Because X-ray telescopes do not have the resolution required to resolve the accretion disc area from the pc-scale jet, this paper is the first to present observational evidence that can pinpoint the point of origin of at least those long-timescale X-ray bursts with durations of 1-3 yrs.

\end{abstract}

\keywords{galaxies: active ---galaxies: jets ---X-rays: bursts}

\section{Introduction}

Variable radio loud active galactic nuclei (AGN) with their associated jets have been studied for more than 40 years. At radio frequencies tremendous advances in their understanding came with the better than 1 milliarcsec resolution of the VLBA which allowed disturbances propelled outward from near the central black hole to be tracked as they moved rapidly outward in a tightly confined jet. Similarly, advances came at X-rays when the $Chandra$ telescope came on-line. Unfortunately, the resolution at X-ray wavelengths is much lower (0.5 arcsec) than with the VLBA. After 40 years of study the exact mechanism by which the jets are formed and the ejected material is accelerated is still unclear. Even the exact point of origin of the non-varying radio component of the AGN core radiation is still being debated, with for e.g. \citet{bel10} presenting arguments favoring a core component centered on the accretion disc and others e.g. \citet{mar10} arguing for a core component located outside the accretion disc area in the base of the jet.

Variations in the total flux of AGN sources at radio frequencies were studied by \citet{and78} when large, single-dish telescopes were first coming on line. Before it was possible to map these sources with the milliarcsec resolution of the VLBA it was demonstrated by \citet[see Fig 1 below]{leg84} that the flux bursts from different AGN sources, though differing slightly in time scale, had similarly shaped profiles. In each event the flux was observed to increase to a peak within $\sim1$ year and then decrease exponentially, falling below the detection limit within 1-3 years. Now, using the VLBA, individual bursts can be tracked at radio frequencies as separate plasma blobs that move away from the central object, reach a peak and then slowly dissipate over 1-3 years \citep[see their Fig 6]{cha11}.

Before a complete understanding of what is happening in the nuclear region of these sources is likely to be achieved it is essential that the exact point of origin of the different radiation components be determined. One of the biggest uncertainties remaining is where the bursting component of the X-ray radiation comes from in AGNs \citep{gal05,bec11}. Is it from material associated with the accretion disc, as argued by \citet{mar06}, or is it from the pc-scale jet? These components cannot be resolved by $Chandra$. Although some X-ray radiation has been detected in the outer, kpc-scale jet and terminal knot of 3C 111 \citep{hog11}, most comes from the unresolved core region containing the accretion disc and pc-scale jet. Much work has been done examining Seyferts and X-ray binaries \citep{zdz03,gal03,kin11}, but in most cases these observations involve variations with shorter timescales (less than 0.3 yr) than those being examined here. Before proceeding, it is important to define exactly what timescale is referred to here when the term "burst" is used. In our case the burst timescale is that defined in early variable source work and covers a period of 1-3 yrs, as defined by a typical radio outburst at cm wavelengths. This has been shown by VLBA observations to be the time taken for the radiation from a new radio outburst to reach a peak and dissipate as it moves out in the pc-scale jet. There are many bursts at all wavelengths that have shorter time scales but this likely means that they are occurring in relatively small structures near the super-massive black hole.

\begin{figure}
\hspace{-1.0cm}
\vspace{-2.0cm}
\epsscale{1.0}
\includegraphics[width=9cm]{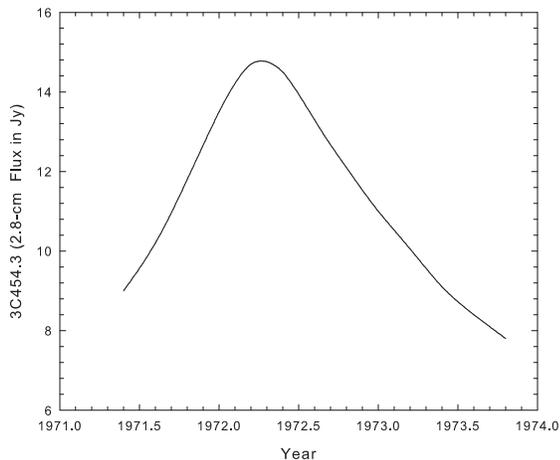}
\caption{{Typical radio burst profile from Legg (1984). \label{fig1}}}
\end{figure}

From Fig 14 of \citet{cha11} the X-ray radiation from 3C 111 can be seen to be composed of two main components of roughly equal peak strengths. These are: 1), a steady component and 2), a superimposed bursting component. The former is assumed to be associated with the accretion disc area. It is this component that may be quenched briefly just prior to the start of each new radio event, producing the dip that was reported by \citet{cha11}. Its appearance close to the event start time requires that it originate close to the nucleus. The latter variable component, is composed of several X-ray bursts that are similar to, or slightly brighter than, the steady component, and it may be possible to determine the point of origin of this component by carrying out cross correlations a) between the radio and X-ray light curves and b) between the radio event start times and the peaks in the X-ray light curve. Although there may be a steady X-ray component associated with the accretion disc, here we are concerned mainly with the bursting one, and we show in Sections 3 and 4 that most of the bursting component of the X-ray continuum in the FR-II radio galaxy 3C 111 originates in the plasma blobs in the pc-scale jet as opposed to the accretion disc area.

\section{The Data}

Recently \citet{cha11} have presented a significant study of the bursting components of the radiation from 3C 111 at several frequencies ranging from 15 GHz to X-rays. This included VLBA maps of the jet structure at 37 GHz for $\sim50$ different epochs between 2004 and 2010. At radio frequencies, with the milliarcsec resolution provided by the VLBA the progression of each newly ejected blob is easily followed as it moves outward in the inner pc-scale jet. Where it was only assumed in the early days that the variable component of the radio flux was associated with the ejected blobs, it can now easily be seen from a series of consecutive VLBA maps that this is the case. Also, 3C 111 is a much better source to use for this purpose than is 3C 120, BL Lac, or PKS 1510-089 whose outbursts are so frequent that their light curves are much more confused.

\citet{cha11} also reported seeing a temporary dip in the X-ray radiation just prior to the ejection of each new radio event. Although not a perfect correlation these dips do appear convincing and may be due to a decrease in the otherwise non-varying X-ray radiation originating near the accretion disc. Although the timing of these dips precedes the estimated event start times by $\sim50$ days, there is no mention of whether any allowance was made for an acceleration period between the actual event time and the first appearance of a new blob. In fact, the dips may occur much closer to the actual event start time. It is assumed here that they may be due to some kind of initial quenching that occurs in the non-varying X-ray component associated with the accretion disc when new material is first injected. The presence of an X-ray dip appearing shortly prior to the ejection of a radio blob was also used by \citet{mar02} to argue for a connection between the black hole and the jet in 3C 120. Unfortunately, as noted above, because of the poor resolution at X-rays, it is not possible to pin-point the point of origin of the X-ray radiation. We can assume, however, that X-ray variations that correlate closely in time (within a few tens of days) with the start of ejection events are likely to be tied to the accretion disc area while peaks in the X-ray light curve that correlate with radio ejection events, but appear a long time after the start of the radio event itself, must originate much further out in the jet.

\begin{figure}
\hspace{-1.0cm}
\vspace{-1.0cm}
\epsscale{1.0}
\includegraphics[width=9cm]{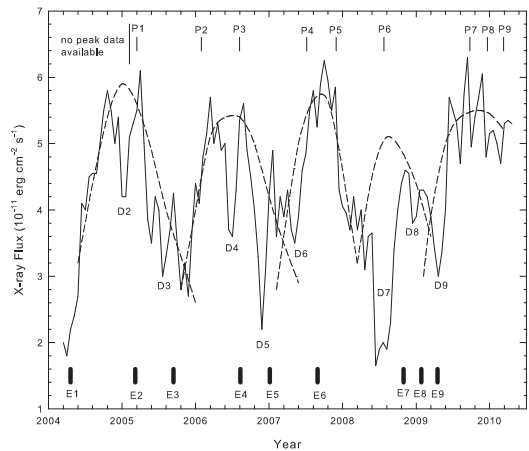}
\caption{{Plot of X-ray flux between 2004 and 2010 from Chatterjee et al.(2011). Dips discussed by Chatterjee are shown shaded and are indicated by D. Times when new events occurred are indicated by E while the times that the flux from these events is estimated to peak, from Fig 1 of Legg (1984), are indicated by P. Estimated burst profiles are indicated by the dashed curves. Events 2 and 3, 4 and 5, and 7, 8, and 9 are considered as single events for the purpose of defining profiles. \label{fig2}}}
\end{figure}

In Fig 2 the solid curve represents the X-ray data obtained for 3C 111 by \citet{cha11} and was reproduced here from a digitized version of their smoothed curve. The times at which each new radio event began are indicated by a vertical bar and the letter E. Also in Fig 2 the dips reported by \citet{cha11} have been highlighted and the dashed curves give a rough indication of how the light curve would be expected to look if the dips were not present. 

When the light curves of 3C 111 obtained at different wavelengths are compared \citep[see their Fig 4]{cha11}, it is apparent that for each burst the shorter wavelengths lead those at radio, with the X-ray radiation leading the radio by $\sim100$ days. Using this information together with the profile shape from Fig 1, if the X-ray radiation comes from the jet, we can predict approximately when each X-ray peak will be expected to occur, at least for each peak that occurs after 2005.0, and these times are indicated in Fig 2 by P. It is immediately apparent that there is good agreement between the predicted and observed X-ray peaks for all cases except during 2008. Here, when E6 is predicted to be peaking, three new events (E7, E8, and E9) are about to begin and the previously reported dips may also be occurring. In addition to this, for a long period (2008.3 to 2008.6) no data are available.

There appears to be little evidence that in each burst the increase in flux and its drop-off again, as outlined by the dashed curves, is related to Doppler boosting, since each blob appears to move out with a constant velocity in Fig 7 of \citet{cha11}.

\begin{figure}
\hspace{-1.0cm}
\vspace{-2.0cm}
\epsscale{0.9}
\includegraphics[width=9cm]{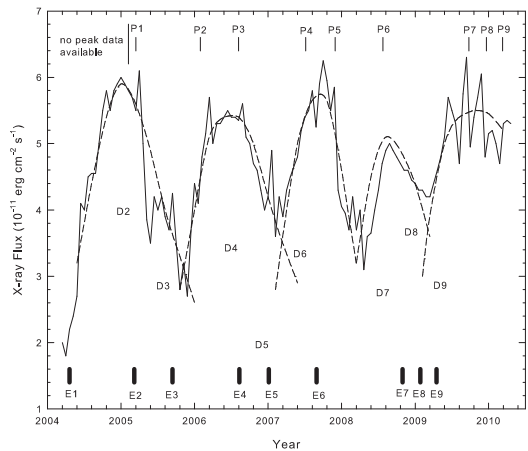}
\caption{{Same as Fig 2 but with dips removed. \label{fig3}}}
\end{figure}

\begin{figure}
\hspace{-1.0cm}
\vspace{-1.0cm}
\epsscale{0.9}
\includegraphics[width=9cm]{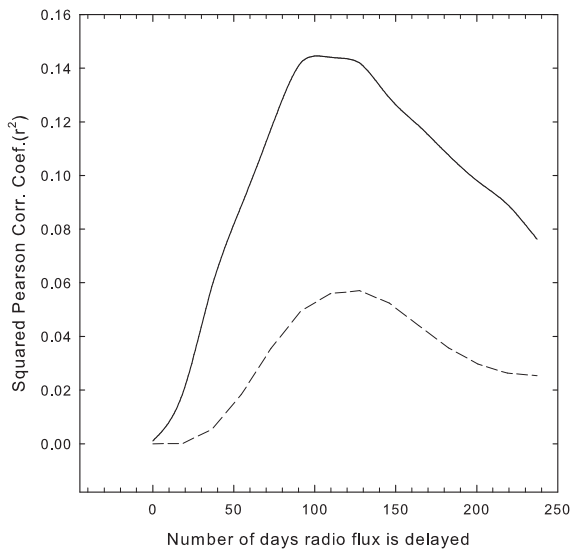}
\caption{{Squared correlation coefficient between X-ray and 14.5 GHz radio light curves plotted as a function of delay in number of days that radio follows the X-ray. (solid curve) data prior to yr 2008.3. (dashed curve) all data. \label{fig4}}}
\end{figure}

\section{Cross-Correlation Between 14.5 GHz Radio and X-ray Light Curves}

In Fig 3 the dips highlighted in Fig 2 have been removed. Since they appear close in time to the new event, they must originate near the accretion disc and will not then be part of the jet radiation. Since their presence could confuse the cross-correlation analysis between the event times and the light curves from the jet that we want to examine, they have been removed prior to the initial calculation. Note that a second analysis using data that included the dips was also carried out to insure that the removal of the dips was not significantly affecting the correlation results.
 
Since the bursting component at radio frequencies is known, from the VLBA maps, to come from the blobs ejected in the pc-scale jet, a strong correlation between the radio (14.5 GHz) and X-ray light curves would be strong evidence that the bursting X-ray radiation also originates in the pc-scale jet. We therefore carried out a cross-correlation analysis between these two curves shown in Fig 4 of \citet{cha11}. To do this the two curves were first digitized into bins of length 0.05yr. The correlation coefficient was then calculated between the two curves as one was shifted relative to the other.

In Fig 4 the squared Pearson correlation coefficient obtained in this analysis is plotted as a function of the delay in days between the radio and X-ray peaks. The solid curve was obtained using only that data in Fig 3 before yr 2008.3. This portion of the data was looked at first because it is difficult to know how the previously mentioned 4-month data gap in yr 2008, and the presence of three consecutive dips (associated with radio events E7, E8, and E9) would affect the results. For the data before 2008.3 a very significant correlation is obtained ($P$ = $2\times10^{-4}$). The result also indicates that the radio peaks are delayed by approximately 100 days. The squared Pearson correlation coefficient obtained using all the data in Fig 3 is shown by the dashed curve in Fig 4. Only a slightly poorer probability ($P$ = $5\times10^{-4}$) is obtained. The correlation analysis was then repeated using the data in Fig 2 which contains the dips. The result was only marginally worse and still resulted in a highly significant probability. This indicates that the strong correlation found has not been produced by the removal of the dips but is, instead, only slightly strengthened, as predicted. This further indicates that the accuracy of the procedure used to remove the dips cannot significantly affect the correlation results.  This gives us strong evidence to argue that the radio and X-ray peaks are truly correlated, which in turn is grounds to argue that the radio event start times may also be correlated with the X-ray peaks if the appropriate time delay is used.

\section{Cross-Correlation Between Radio Event Start Times and Peaks in the X-ray Light Curve}

We then carried out a cross-correlation analysis between the start times of the radio events and the X-ray light curve. The results are shown by the solid curve in Fig 5. Here the correlation coefficient has been obtained between the radio event start times and the structure in the X-ray light curve in Fig 3, as one curve is shifted relative to the other. As predicted from the curve in Fig 1, a correlation peak is seen between the radio event start times and the peaks in the X-ray light curve that occurs when the radio curve is shifted by 328 days relative to the X-ray light curve. This corresponds to a shift of 17.9 channels, where the channel width is set by the bin size of 0.05 yr (18.3 days) used in the digitization process. The probability represented by this peak is $P$ = 0.013 and is quite significant. But this should be no surprise since it is already obvious from Figs 2 and 3 that the predicted P-values align well with the peaks in the X-ray light curve. Although there is a weaker correlation peak obtained when the radio and X-ray curves are shifted relative to each other by channel shifts between 2 to 6 in Fig 5, it is negative in $r$, where $r$ is the correlation coefficient, and only becomes positive when $r$ is squared. This peak represents an anti-correlation associated with the valleys between the peaks. A cross-correlation was also carried out between the radio event start times and the data in Fig 2, which still contains the X-ray dips. Again it showed only a marginal worsening in the correlation coefficient when the dips were included. Unlike the negative dip components, which must originate in material near the accretion disc, there is now little doubt that the positive, bursting X-ray components arise in the plasma blobs in the pc-scale jet, well outside the accretion disc.

\begin{figure}
\hspace{-1.0cm}
\vspace{-2.0cm}
\epsscale{0.9}
\includegraphics[width=9cm]{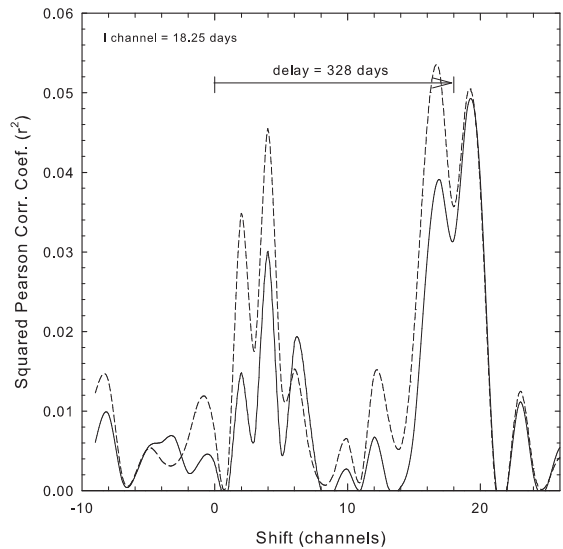}
\caption{{Squared correlation coefficient for data in Fig 3. Event times are shifted and correlated with the X-ray light curve. The peak between 2 and 6 channels is anti-correlated (negative) but inverted when the correlation coefficient is squared. It is associated with the valleys between the bursts. See text for an explanation of the dashed curve. \label{fig5}}}
\end{figure}
 
\begin{figure}
\hspace{-1.0cm}
\vspace{-2.0cm}
\epsscale{0.9}
\includegraphics[width=9cm]{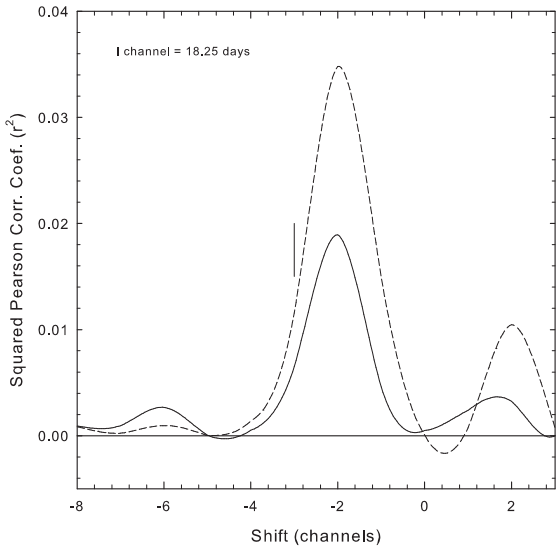}
\caption{{(solid curve)Squared correlation coefficient for data in Fig 2. Event times are shifted and correlated with the X-ray light curve. Here the peak is negative prior to squaring because the dip is negative. Vertical bar indicates location of the correlation peak found by Chatterjee with dips 6 and 9 omitted from the calculation. (dashed curve) Same for data in Fig 2 with low weight given to event 6.\label{fig6}}}
\end{figure}

\begin{figure}
\hspace{-1.0cm}
\vspace{-0.5cm}
\epsscale{0.9}
\includegraphics[width=9cm]{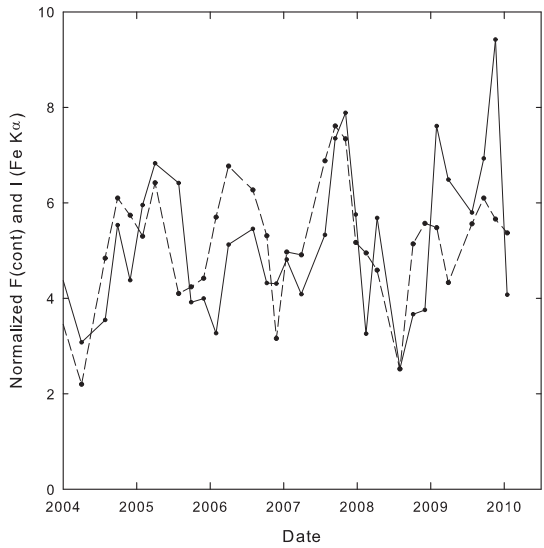}
\caption{{Fe K$\alpha$ line (solid) and F(2-10) continuum (dashed) plotted versus date. Each peak can be seen to rise first in the continuum. Data are from Chatterjee (2011). \label{fig7}}}
\end{figure}

\begin{figure}[ht]
\hspace{-1.0cm}
\vspace{-1.0cm}
\epsscale{0.9}
\includegraphics[width=9cm]{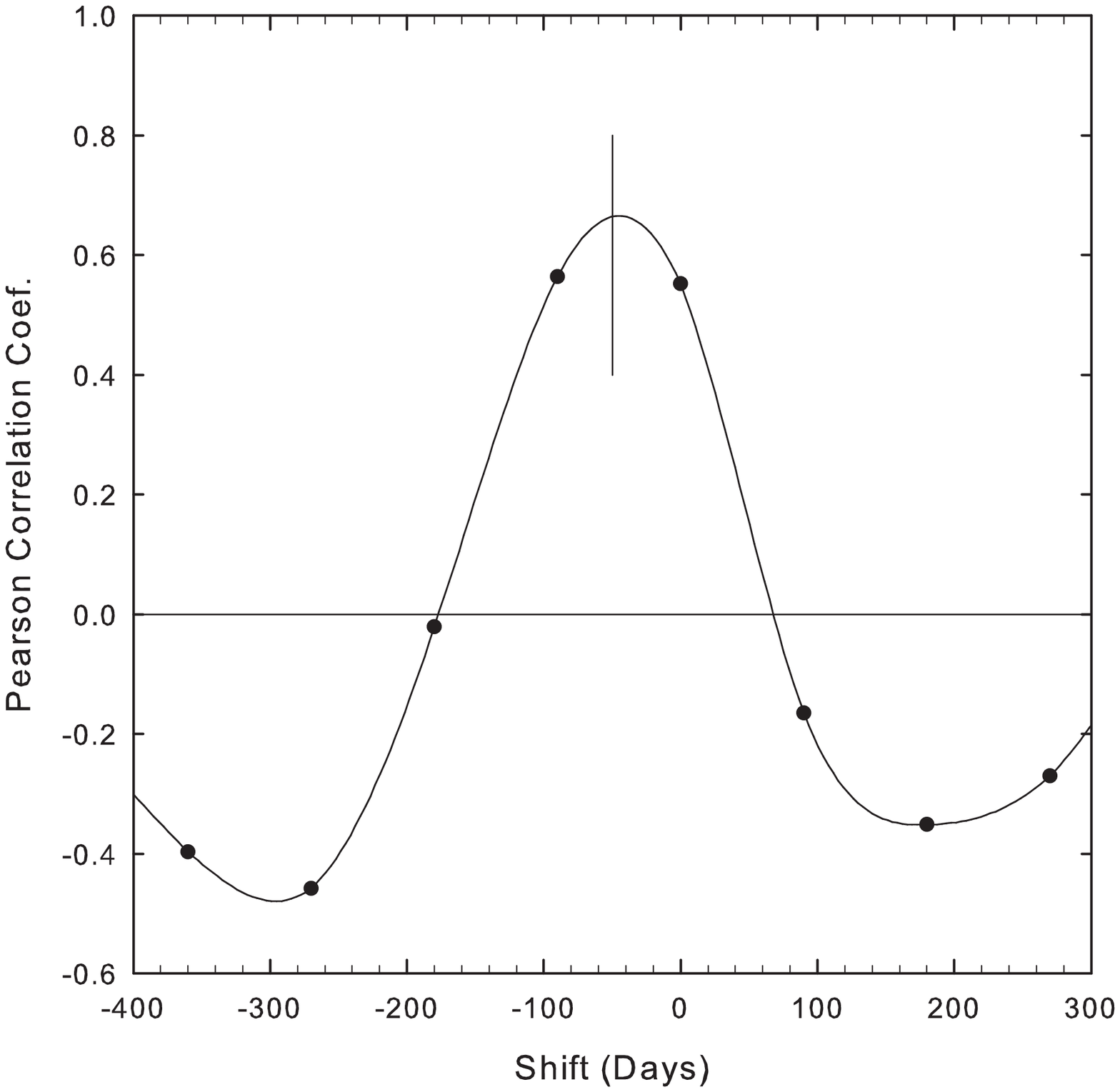}
\caption{{Pearson correlation coefficient plotted versus shift of Fe K$\alpha$ line relative to F(2-10) continuum. The curve center indicates that the Fe K$\alpha$ line variations lag $\sim$50 days behind those of the continuum. \label{fig8}}}
\end{figure}

\section{Discussion}

Although there is no evidence for a correlation peak in Fig 5 near -2 channels, which would be produced by the dips reported by \citet{cha11}, this is expected since the analysis has been carried out on the data in Fig 3 where the dips have been removed. In Fig 6 a cross-correlation between event times and the data in Fig 2, which still contains the dips, is given by the solid curve. There is now evidence for a peak, although with a probability of $P$ = 0.061 it is only marginally significant. It is found to occur $\sim36$ days prior to the event start times where \citet{cha11} found a value of 54 days. This may be due to the fact that these authors did not include events 6 and 9 in their calculations. In fact, there seems to be some question about the reality of burst E6 which, in Fig 7 of \citet{cha11}, appears to be confused with E5. 

 To examine the effect that events 6 and 9 might have on the correlation results, the correlation was re-run giving each of these events low weight in turn. It made no difference to the result in Fig 6 whether event 9 was included in the calculation or left out. However, when event 6 was given low weight the significance was greatly enhanced resulting in a correlation coefficient of $r^{2}$ = 0.035 and a probability of $P$ = 0.019, as is shown by the dashed curve in Fig. 6. However, it did not change the time found between the dip and its associated event.

Giving burst E6 low weight in the cross-correlation calculation in Fig 5 resulted in only a marginal increase in significance as is shown by the dashed curve in Fig 5. 


X-rays have been found previously to be associated with radio blobs in the kpc-scale jet of 3C 111 and other sources where the blobs can be resolved by $Chandra$ as well as the VLBA \citep{hog11,kat03}. That work has established that, although the radiation from the core region always dominates, X-ray emission from large scale jets of radio galaxies and quasars is more common than expected with X-ray jets detected in 78$\%$ of the sources studied.

The results found here contradict what is usually assumed to be the case, namely that the bursting X-ray component is associated with the accretion disc area \citep{fab00}. However, this may not be too surprising since others \citep{sul98,sue98} have concluded that the evidence for and against an accretion disc origin for the Fe K$\alpha$ emission is equivocal.

\citet{cha11} have reported that the fluctuations in the X-ray continuum and Fe K$\alpha$ line are correlated. In their study these parameters were sampled approximately every 0.25 yr between 2004 and 2010. In order to examine the variations in these parameters more closely we have plotted them here in Fig 7. To make the comparison easier the Fe K$\alpha$ line values have been multiplied by a factor of 0.6 to normalize the two curves.  Both the (2-6 Kev) continuum and the Fe K$\alpha$ line intensity clearly show the same peaks. Furthermore, these are the same X-ray peaks found above to be correlated with the radio peaks and, with a long delay, with the event start times. In each peak in Fig 7 the continuum intensity rises slightly before that of the Fe K$\alpha$ line. A cross-correlation between these two lines is shown in Fig 8. As found by \citet{cha11} this peak represents a significant correlation with a probability of $P$ = $4\times10^{-4}$. Fig 8 indicates that the Fe line lags the continuum by $\sim50$ days and it can be assumed that the variable portion of the line flux originates in material located within $\sim50$ lt-days of the (2-6 Kev) continuum radiation source.

If the X-ray continuum radiation originates in the pc-scale jet, as found above, this evidence then indicates that the Fe K$\alpha$ line radiation must originate nearby, and from the excellent correlation found it seems likely that the Fe K$\alpha$ line is somehow associated with the jet. It cannot originate inside the jet if the jet is moving relativistically, since the Fe line is not significantly shifted. It may originate, however, from a bow wave produced by the ejected "blob" as it plows through an out-flowing wind. This might be expected to produce some line broadening, but little line shift, as seen, since the speed of any outflow or wind will be quite low compared to that of the material in the jet. That the Fe K$\alpha$ line radiation would be correlated with the appearance and disappearance of the continuum bursts as seen, would also be expected in this scenario.

To the best of our knowledge there is no bursting radiation associated with the kpc-scale components that would have been significant in the single-dish radio measurements. Therefore the bursting radiation must be associated either with the accretion disc or the pc-scale jet. Because the X-ray telescope does not have the resolution required to resolve the accretion disc area from the pc-scale jet, this paper is the first to present observational evidence that can pinpoint the point of origin of the bursting X-ray component. 

Although the purpose of this study is to determine exactly where the bursting component of the X-ray radiation originates, we assume that the X-ray emission mechanism in the pc-scale jet is similar to that suggested for the external jets of powerful FR II sources which has been interpreted as inverse Compton scattering off of cosmic microwave background photons by the electrons in the relativistic jets. Recently, more and more evidence is being found showing that there is a connection between cm/mm radio and high energy emission in quasars, especially gamma-ray emission which has also been shown to come from the pc-scale jet. For example, \citet{sch11} have recently shown that the gamma-ray emission in 3C 345 is produced, not in a compact region near the central engine of the AGN, but in the Compton-loss dominated zone of the pc-scale jet. It should be noted that although these authors refer to 3C 345 as the archetypal quasar, this source has also been classified as an Ultra Luminous Infrared Galaxy by \citet{hou09}.

\section{Conclusions}

Using a cross-correlation analysis we have found a strong correlation between the 14.5 GHz radio and the X-ray light curves of 3C 111, with the radio peaks trailing the X-ray by $\sim100$ days. We have also demonstrated that this bursting component of the X-ray radiation originates, as does the bursting component of the radio radiation, in the ejected plasma blobs as they move outward in the pc-scale jet of 3C 111, peaking $\sim1$ year after the start of each new event. The fact that a strong correlation exists between the start of each new radio event and its associated X-ray peak that occurs $\sim328$ days later, indicates that the X-ray radiation must be associated with the radio blobs in the pc-scale jet that peak on this same time scale. Since the radio peak occurs $\sim100$ days after the X-ray peak, it will occur $\sim428$ days (14.3 months) after the beginning of each new event. This is in good agreement with what was found for the radio bursts by \citet{leg84}. A correlation was also found here between the event start times and the X-ray dips that precede them, supporting the previous claims by \citet{cha11}.

\end{document}